\documentclass[twocolumn,prl,tightenlines,superscriptaddress,showpacs]{revtex4-1}

\usepackage{amsmath}
\usepackage{amssymb,amsfonts,latexsym}
\usepackage{bm}
\usepackage[mathcal]{euscript}
\usepackage{graphicx}
\usepackage{epsfig}
\usepackage{color}

\begin{document}

\title{Understanding dense active nematics from microscopic models}

\author{Aurelio Patelli}
\affiliation{Service de Physique de l'Etat Condens\'e, CEA, CNRS, Universit\'e Paris-Saclay, CEA-Saclay, 91191 Gif-sur-Yvette, France}

\author{Ilyas Djafer-Cherif}
\affiliation{Service de Physique de l'Etat Condens\'e, CEA, CNRS, Universit\'e Paris-Saclay, CEA-Saclay, 91191 Gif-sur-Yvette, France}

\author{Igor S. Aranson}
\affiliation{Department of Biomedical Engineering, Pennsylvania State University, University Park, PA 16802}

\author{Eric Bertin}
\affiliation{LIPHY, Universit\'e Grenoble Alpes and CNRS, 38000 Grenoble, France}

\author{Hugues Chat\'{e}}
\affiliation{Service de Physique de l'Etat Condens\'e, CEA, CNRS, Universit\'e Paris-Saclay, CEA-Saclay, 91191 Gif-sur-Yvette, France}
\affiliation{Computational Science Research Center, Beijing 100094, China}
\affiliation{LPTMC, Sorbonne Universit\'e, CNRS, 75005 Paris, France}

\date{\today}

\begin{abstract}
We study dry, dense active nematics at both particle and continuous levels.
Specifically, extending the Boltzmann-Ginzburg-Landau approach,
we derive well-behaved hydrodynamic equations from a Vicsek-style model with nematic alignment and pairwise repulsion.
An extensive study of the phase diagram
shows qualitative agreement between the two levels of description. We find in particular that
the dynamics of topological defects strongly depends on parameters and can lead
to ``arch'' solutions forming a globally polar, smectic arrangement of N\'eel walls. We show how these configurations
are at the origin of the defect ordered states reported previously.
This work offers a detailed understanding of the theoretical description of dense active nematics 
directly rooted in their microscopic dynamics.
\end{abstract}

\maketitle

Active nematics (collections of self-propelled elongated particles aligning by collisions) 
has been the subject of rather intense theoretical attention
\cite{YEOMANS-EARLY,Giomi-EARLY,SHI-MA-NAT-COMM,YEOMANS-DEFECT,MCM-DEFECT1,MCM-DEFECT2,YEOMANS-FRICTION,YEOMANS-VORTICITY,GIOMI-PRX,WET-TO-DRY,INTRINSIC,SRIVASTAVA,BASKARAN,YEOMANS-CELL-DEFECT,HEMINGWAY,YEOMANS-JFM,DUNKEL,ACTNEMA-NATREVIEW,SHANKAR,LIVERPOOL18,SELINGER19}. 
This flurry of papers was largely triggered by a few remarkable experiments 
\cite{DOGIC-NATMAT,ZHOU,CREPPY,SAGUES,SAGUES2,DUCLOS,KAWAGUCHI,LADOUX,GENKIN17,GENKIN18} ---see also \cite{DOGIC17,ARANSON18} for recent reviews.
In particular, Dogic {\it et al.} \cite{DOGIC-NATMAT} studied a suspension of bundles of stabilized microtubules and clusters of kinesins motor proteins
sandwiched at an oil-water interface. This revealed sustained regimes of `nematic turbulence' with prominent 
motion of $\pm\frac{1}{2}$ topological defects, including intriguing `defect ordered states' in which $+\frac{1}{2}$ defects
seem to be ordered on very large scales despite an overall disordered background.

Most theoretical efforts towards accounting for these remarkable experiments fall into two categories: 
microscopic, active particle models \cite{SHI-MA-NAT-COMM,DOGIC-NATMAT,LI-PNAS} 
and continuum descriptions in terms of deterministic hydrodynamic equations
\cite{MCM-DEFECT1,YEOMANS-VORTICITY,GIOMI-PRX,INTRINSIC}.
The microscopic models are mostly `dry', i.e., they neglect the fluid surrounding the active particles, while mostly `wet' continuous descriptions,
{\it i.e.} including a Stokes equation describing this fluid, have been considered.
Dry continuous descriptions
have also been written by enslaving the fluid 
\cite{BASKARAN,WET-TO-DRY,SRIVASTAVA}.
Whereas all these works have had some success in accounting for the properties observed experimentally, 
they lack a direct connection between the microscopic and the macroscopic level: the proposed continuous descriptions have not been 
studied in parallel to particle-based models.
It is thus impossible to relate clearly their many parameters to those of any underlying microscopic dynamics.

In this Letter, we bridge the gap between particle-based models and continuous 
theories for dense active nematics systems, restricting ourselves to the dry case. 
Extending the Boltzmann-Ginzburg-Landau approach of \cite{BGL,RODS,NEMAMESO},
we derive well-behaved hydrodynamic equations from a Vicsek-style active nematics model with alignment and repulsion interactions. 
An extensive study of the phase diagrams of both
the particle model and the hydrodynamic equations (varying the key parameters of the microscopic dynamics) shows
qualitative agreement between the two levels of description. 
Among our salient results, we find that the dynamics of topological defects strongly depends on specific parameters. 
We also uncover `arch' solutions forming a globally polar, smectic arrangement of N\'eel walls that coexist with the homogeneous nematic state
in large regions of parameter space. We show them to be at the origin of `defect ordered states' reported previously in \cite{DOGIC-NATMAT,BASKARAN}.
Our work offers an understanding of the theoretical description of dense active nematics 
directly rooted in their microscopic dynamics and a unified account of previous partial results.

The microscopic, dry models proposed so far for dense active nematics rely on volume exclusion effects between 
elongated particles \cite{SHI-MA-NAT-COMM, DOGIC-NATMAT}. 
Here we use a Vicsek-style model instead, where pointwise particles interact with those
within a fixed distance \cite{VICSEK,CHATE}.
This makes both the derivation of hydrodynamic equations and an 
extensive numerical study easier. Specifically, we study a variant of the model introduced in \cite{SMECTICS} for the study of active smectics. 
We consider point particles moving at constant speed $v_0$ along the unit vector 
${\bf e}(\theta)$ defined by their heading $\theta$
in a rectangular domain of size 
$L_x \times L_y$ with periodic boundary conditions. 
At discrete unit timesteps, the position ${\bf r}_i$ and heading $\theta_i$ of particle $i$ 
are updated according to: 
\begin{eqnarray}
{\bf r}_i^{t+1}&=&{\bf r}_i^t + v_0\, {\bf e}(\theta_i^{t+1}) \\
 \theta_i^{t+1} &=& \text{arg}\left[ \epsilon(t) \langle \text{sgn}(\cos(\theta_i^t - \theta_j^t))\,{\bf e}(\theta_j^t)\rangle_j \right.  \nonumber \\
&&\left. + \beta \langle{\bf \hat r}_{ji}^t \rangle_{j} \right] + \eta \chi_i^t
\label{eq:angle_update}
\end{eqnarray}
where $\chi_i^t\in[-\frac{\pi}{2}, \frac{\pi}{2}]$ is an angular white noise drawn from a uniform distribution, 
$\eta$  is a parameter setting the strength of the angular noise, 
$\epsilon=\pm1$ reverses sign with probability rate $k\in[0,0.5]$, 
${\bf \hat r}_{ji}$ is the unit vector pointing from particle $j$ to $i$,
and the average is taken over the neighbors $j$ within unit distance of particle $i$ 
(including $i$ for the alignment, i.e. the first term in \eqref{eq:angle_update}).
The pairwise repulsion plays the role of a `torque' and has constant modulus with coupling $\beta$.
The interaction range is the same for both alignment and repulsion.

\begin{figure}[t!]
\includegraphics[width=\columnwidth,clip=on]{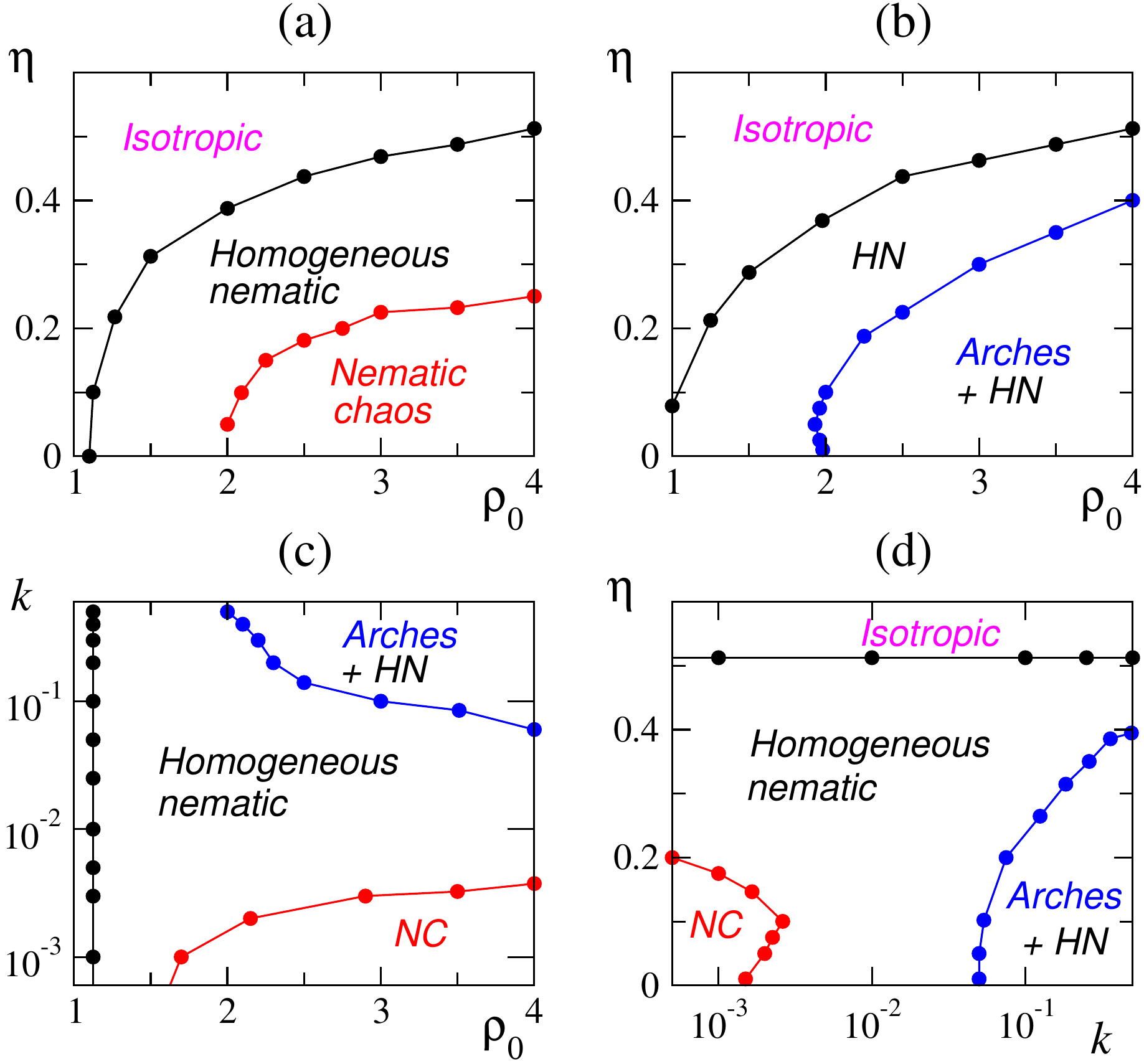}
\caption{(color online) Phase diagram of the particle model 
($\beta=0.5$, $v_0=0.3$, $256\times 256$ periodic box).
(a,b): $(\rho_0,\eta)$ plane for the rods ($k=0$) and active nematics ($k=0.5$) cases.
(c): $(\rho_0,k)$ plane at $\eta=0.1$.
(d): $(k,\eta)$ plane at $\rho_0=4$.
}
\label{fig:micro-phasediagram}
\end{figure}

Without repulsion ($\beta=0$), 
this model, for large reversal rate $k$,  is the minimal model for active nematics introduced in \cite{CGM} and 
studied more recently in \cite{NEMANEW,PERUANI}, while without velocity reversals ($k=0$) it is the Vicsek-style model
for `self-propelled rods' of \cite{RODS}. 
Figures~\ref{fig:micro-phasediagram}a,b show how the typical $(\rho_0,\eta)$ phase diagram of these repulsion-free models 
changes in the presence of repulsion. 
First, the band-chaos coexistence phase bordering the order/disorder black line is too small to be seen 
at the system sizes and densities involved. 
Repulsion induces the emergence of new phases below this black line:
For $k=0.5$ (Fig.~\ref{fig:micro-phasediagram}b), ``arch'' solutions (described in detail below) {\it coexist} with the defect-free nematic liquid at large-enough density (blue region).
For rods ($k=0$, Fig.~\ref{fig:micro-phasediagram}a), an {\it inner} region of nematic chaos (in red) appears deep in the ordered phase
\footnote{At low densities, a small region of {\it polar} bands with local smectic order is present
that quickly disappears for non-zero reversal rates. 
These bands constitute a marginal phenomenon in the context of this paper.}.
Figures~\ref{fig:micro-phasediagram}c,d show how the two limit cases above ($k=0$ and $k=0.5$) are connected when the reversal rate $k$ is varied.
In the $(\rho_0,k)$ plane (at $\eta=0.1$, Fig.~\ref{fig:micro-phasediagram}c), 
the regions of nematic chaos and arch solutions get closer to each other as $\rho_0$ gets larger. 
The gap in between becomes narrower as the system size in increased (not shown).
In the $(k,\eta)$ plane (at $\rho_0=4$, Fig.~\ref{fig:micro-phasediagram}d), both the nematic chaos region and the arch solutions 
disappear when approaching the basic isotropic/nematic black transition line. 

Nematic chaos is characterized by {\it local} nematic order but global disorder `mediated' by $\pm\frac{1}{2}$ topological defects.
It has long but finite correlations lengths and times, and is the result of a longitudinal instability of the homogeneous
nematic, similar to that observed generically in the wet case. 
This instability never saturates into regular undulations,
except in small systems. Instead, defects are always nucleated and nematic chaos sets in.

\begin{figure}[t!]
\includegraphics[width=0.46\columnwidth,clip=on]{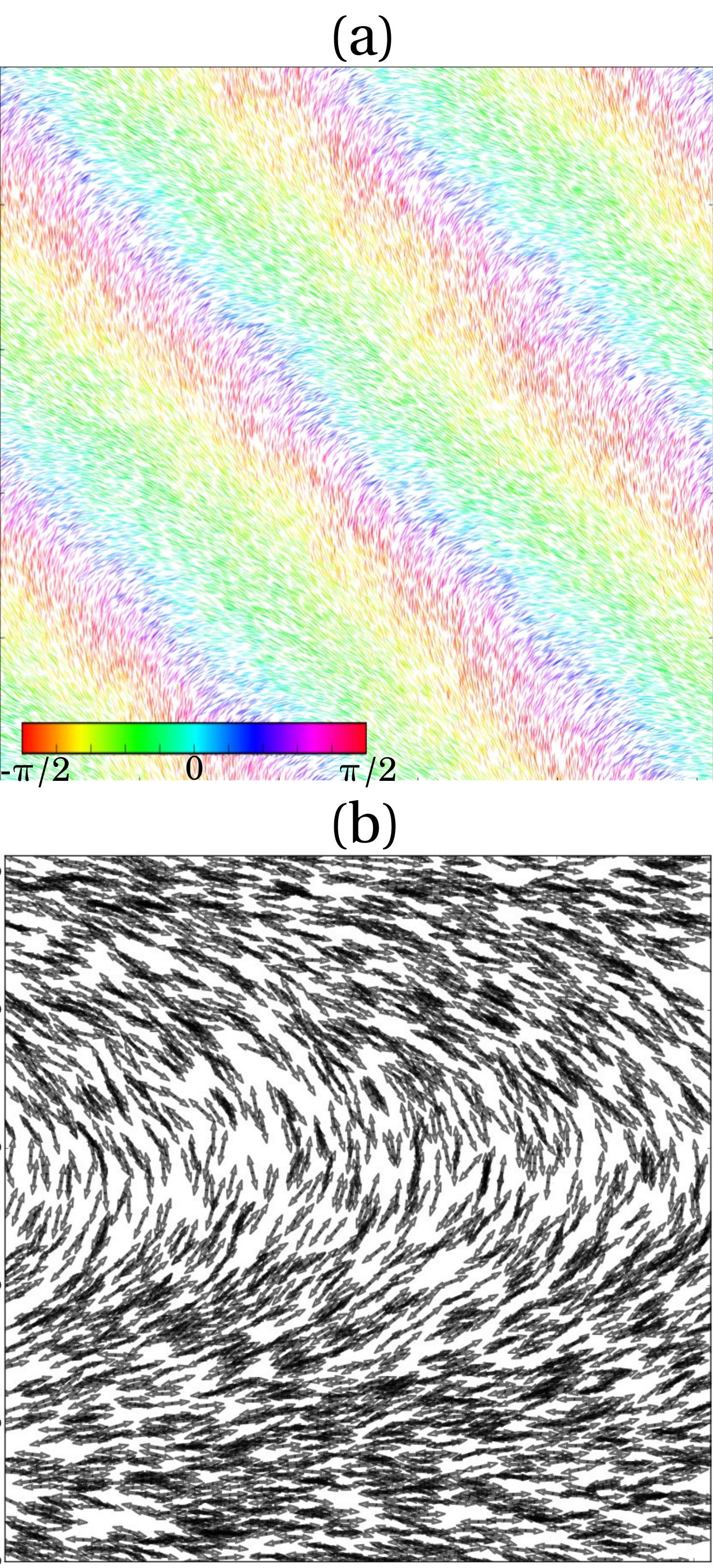}
\includegraphics[width=0.5\columnwidth,clip=on]{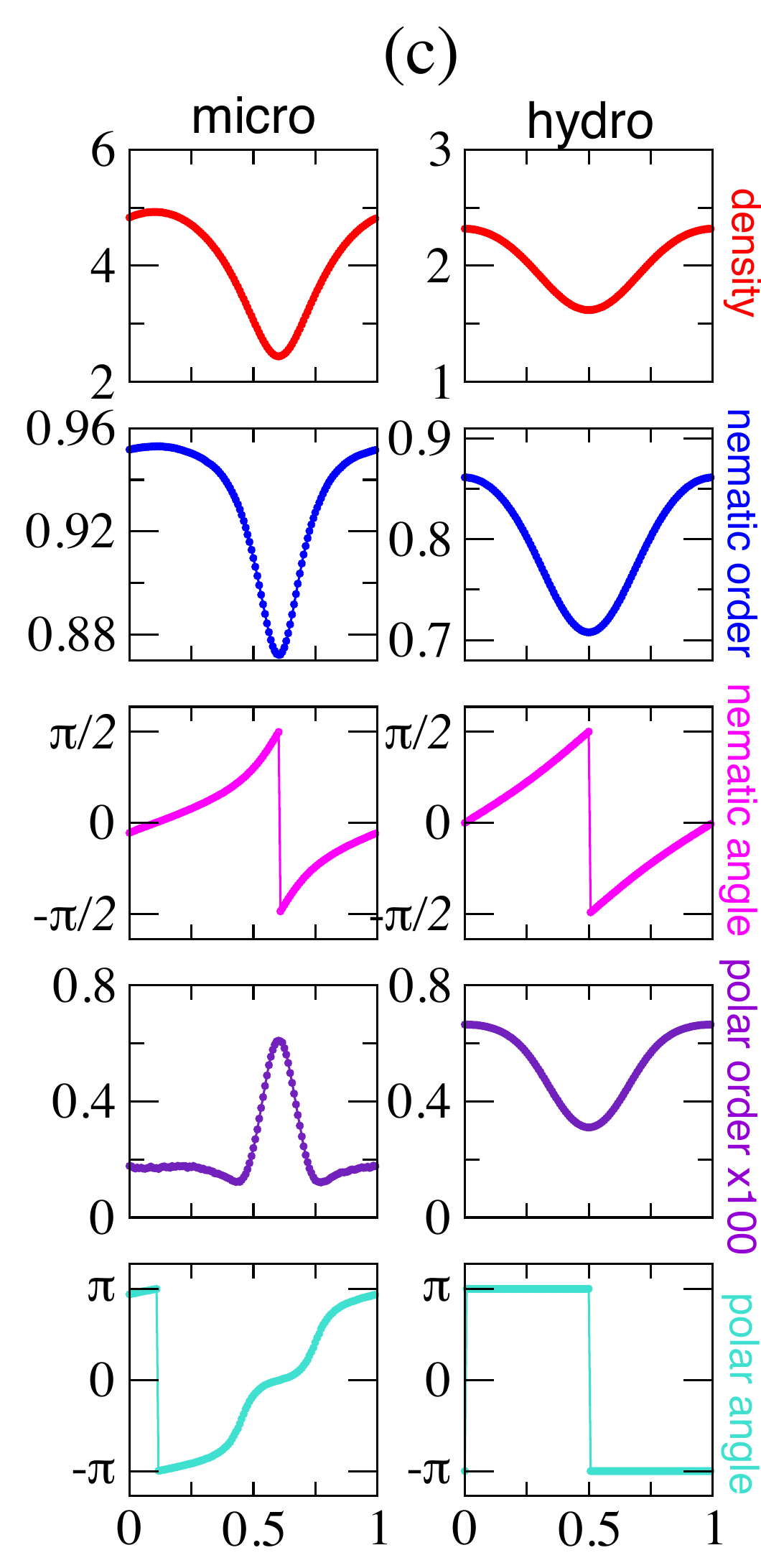}
\caption{Arch solutions.
(a) snapshot of a steady pattern of arches in the microscopic model 
($L=256$, $\rho_0=3$ $v=0.3$, $\eta=0.1$, $\beta=0.5$, $k=0.5$, particles colored by their orientation).
(b) zoom on one of the arches seen in (a) (particles' orientation marked by small double arrows)
(c) profiles of density, nematic and polar order (modulus and direction) 
for the microscopic model (left, same parameters as (a)) 
and the hydrodynamic equations ($\rho_0=1$, $a=10$, $\sigma=0.1$, $b_1=0.25$).}
\label{fig:arches}
\end{figure}

The arch solutions consist of a smectic pattern of
rather sharp walls akin to N\'eel- or $\pi$-walls observed in some equilibrium liquid crystals
 (Fig.~\ref{fig:arches}) \cite{NEEL}. Starting from random initial conditions, local nematic order quickly arises, 
defects are created, and move across the periodic boundaries. In a large-enough system,
this dynamics often results in some solution comprising several arches, which are 
found {\it more frequently} than the defect-free homogeneous nematic state
(Fig.~\ref{fig:arches}a,b). 
Each arch displays local nematic order but is a {\it globally polar} object. Particle trajectories
show weak but regular drift, with a velocity depending on their position across the arch pattern. 
Characteristic profiles of density, nematic 
and drift velocity (polar order) are shown in Fig.~\ref{fig:arches}c. 
Arches do {\it not} have a preferred size: at fixed parameter values,
there is a minimal width below which they disappear, but there is no maximal width (not shown).
They do form a regular pattern, though: starting from a configuration made of $n$ arches of various width,
one observes that the total system slowly evolves toward $n$ arches of equal width. 

We now turn to the derivation of hydrodynamic equations for dense active nematics 
following the approach used for dilute Vicsek-style models in \cite{RODS-KINETIC,NEMAMESO,BGL}.
The starting point is a Boltzmann equation governing $f({\bf r}, \theta,t)$, the probability (density) of finding a particle at 
position ${\bf r}$, with orientation $\theta$, at time $t$.
This equation extends that used in \cite{RODS-KINETIC} with a term describing velocity reversals at rate $a$, 
as in \cite{PERUANI}:
\begin{eqnarray} \nonumber
\partial_t f + v_0 \,\mathbf{e}(\theta)\cdot \nabla\! f  &=&
\lambda [\langle f(\theta-\eta)\rangle_{\eta} -f(\theta)] + I_{\rm col}[f]\\
&& \qquad + a [f(\theta+\pi) - f(\theta)]
\label{eq:Boltzmann-real}
\end{eqnarray}
where $\lambda$ is a tumbling rate, $\nabla$ the gradient operator, and $\langle \dots \rangle_{\eta}$ indicates 
an average over the noise distribution $P(\eta)$. The collision integral 
\begin{eqnarray}
	\label{eq:Collision_Integral-real}
&& I_{\rm col}[f] = \!\! \int_{-\pi}^{\pi}  \!\!\!\! d\theta_1 \!\! \int_{-\pi}^{\pi} \!\!\!\! d\theta_2 f(\mathbf{r},\theta_1) \!\! \int_0^{\infty} \!\!\!\! ds\, s 
	\!\! \int_{-\pi}^{\pi} \!\!\!\!d\phi \, K(s,\phi,\theta_1,\theta_2) \nonumber \\
&& \;\; \times f(\mathbf{r}\!+\!s\mathbf{e}(\phi),\theta_2)\Big[\langle \hat{\delta}\big(\Psi(\theta_1,\theta_2)\!+\!\eta \!-\!\theta\big)\rangle_{\eta} \!-\! 
\hat{\delta}(\theta_1\!-\!\theta)\Big] 
\end{eqnarray}
where $\Psi$ is the $\pi$-periodic nematic alignment function
[$\Psi(\theta_1,\theta_2)=\frac{1}{2}(\theta_2-\theta_1)$ for $-\frac{\pi}{2} < \theta_2-\theta_1< \frac{\pi}{2}$], 
incorporates the distance-dependent repulsive interaction via the dependence of the
collision kernel $\mathrm{K}(s,\phi,\theta_1,\theta_2)$ on the relative position
$\mathbf{r'}\!-\!\mathbf{r} \equiv s\mathbf{e}(\phi)$ 
of the two colliding particles; 
$\hat{\delta}$ is a $2\pi$-periodic Dirac distribution.
Suppose that the two
particles are hard disks of diameter $d_0$. The angle $\phi$ is an impact parameter
defined by the position of the contact point at collision.
The probability to collide within an infinitesimal interval $d\phi$ around $\phi$ is proportional to $\cos(\phi-\theta_{12}) d\phi$ where
$\theta_{12}$ is the direction of $\mathbf{e}(\theta_1)-\mathbf{e}(\theta_2)$ 
[collision occurs only if $\cos(\phi-\theta_{12})>0$]. 
Kernel $K$ is thus modified from its form
$K(\theta_1,\theta_2) =  2 |\sin\frac{\theta_1-\theta_2}{2}|$ for ballistic particles to
\begin{equation}
	K = 2 g(s)\,\left|\sin\frac{\theta_1 \!-\! \theta_2}{2}\right|
		 \cos(\phi \!-\! \theta_{12})\, \Theta\left[\cos\left(\phi \!-\! \theta_{12}\right)\right]
	\label{eq:CollisionKernel_general}
\end{equation}
where $\Theta(x)$ is the Heaviside function and 
$g(s)$ is an integrable function over the interval $[0,d_0]$ modeling repulsion between soft spheres
($g(s)=\delta(s-d_0)$ in the hard spheres limit).
Finally, only small variations of $f$ are expected over the distance $s$ between colliding particles, so that we write
\begin{equation}
\label{eq:grad-exp}
f(\mathbf{r} \!+\! s\mathbf{e}(\phi),\theta_2) \!\approx\! \left[ 1 \!+\! s\mathbf{e}(\phi) \!\cdot\! \nabla \right]\! f(\mathbf{r},\theta_2)
\end{equation}
The tumbling rate $\lambda$ and the elementary size $d_0$ can be set to unity without loss of generality, but the 
particle speed $v_0$ cannot be scaled out here.
Using ({\ref{eq:grad-exp}), the Boltzmann equation is expressed as a hierarchy of equations
in terms of complex modes $f_k$ given by
$f_k(\mathbf{r}) = \int_{-\pi}^\pi d\theta e^{\imath k \theta} f(\mathbf{r},\theta)$. 
In addition to $f_0=\rho$, the density field, the fields of interest here are both $f_1$ and $f_2$ which are related to
polarity vector 
and the nematic tensorial fields
\footnote{The polarization vector is $\mathbf{P}=(\Re\{f_1\},\Im\{f_1\})$ 
while the independent components of the nematic tensor are $\mathbf{Q}_{xx}=\Re\{f_2\}$ and $\mathbf{Q}_{xy}=\Im\{f_2\}$.}.
This hierarchy is truncated and closed 
with the scaling ansatz used in \cite{RODS-KINETIC} for the dilute case, 
$\nabla\sim\partial_t\sim\delta\rho\sim\epsilon$, $f_{2k-1}\sim f_{2k}\sim \epsilon^{|k|}$.
At order $\mathrm{O}(\epsilon^3)$, 
we obtain the closed equations
\begin{eqnarray} 
\label{eq:f0}
	\partial_t \rho &=& -\tfrac{1}{2}v_0\left( \triangledown^*\! f_1 + \triangledown\! f_1^* \right) \\
\label{eq:f1}
	\partial_t f_1&=&(\alpha[\rho]- \beta\vert f_2\vert^2 )f_1  +\zeta f_1^*f_2
	-\pi_0[\rho]\triangledown\! \rho \nonumber \\
	&-&\pi_2[\rho]\triangledown^*\! f_2
	+\gamma_2 f_2\triangledown\! f_2^* +\gamma_1 f_2^*\triangledown\! f_2
	-\lambda_n f_2\triangledown^*\! \rho\nonumber\\
	&+&\lambda_1 f_1\triangledown^*\! f_1+\lambda_2 f_1\triangledown\! f_1^*+\lambda_3 f_1^*\triangledown\! f_1 \\
\label{eq:f2}
	\partial_t f_2 &=&(\mu[\rho] + \tau \vert f_1\vert^2-\xi\vert f_2\vert^2)f_2 + \omega f_1^2+\nu\triangle f_2 \nonumber\\
&-&\pi_1[\rho] \triangledown\! f_1 + \chi_1 \triangledown^*\! (f_1f_2) +\chi_2 f_2\triangledown^*\! f_1 + \chi_3 f_2 \triangledown\! f_1^*\nonumber\\
	&+& \kappa_1 f_1^* \triangledown\! f_2 +\kappa_2 f_1\triangledown\! \rho
\end{eqnarray}
where $\triangledown=\partial_x+i\partial_y$, $\triangledown^*=\partial_x -i\partial_y$, $\triangle=\triangledown\triangledown^*$. 
These equations include those derived in \cite{RODS-KINETIC,BGL,bertin2015} for the case of dilute rods;  
the last line in (\ref{eq:f1}) and (\ref{eq:f2}) contain the new terms due to repulsion, 
whose coefficients depend on the first moments of $g(s)$. 
The expressions of all coefficients have been derived;
only their dependence on $\rho$ is indicated explicitly above.
These equations are formally very close to those obtained when enslaving the fluid in wet active nematics 
\cite{BASKARAN,WET-TO-DRY,SRIVASTAVA}.
A detailed discussion of their structure will be given elsewhere \cite{TBP}. 
They possess two main uniform solutions, the disordered ($\rho=\rho_0$, $f_1=f_2=0$) and 
the nematically-ordered one ($\rho=\rho_0$, $f_1=0$, $|f_2|^2=\mu[\rho_0]/\xi$).
The uniform disorder solution is linearly stable whenever $\mu[\rho_0]<0$. For $\mu[\rho_0]>0$, the nematic uniform solution exists.
Its full linear stability analysis was
performed semi-numerically. 
Without repulsion, and for $a=0$ and $a\to \infty$, we recover the simple phase diagrams found respectively in \cite{RODS-KINETIC,NEMAMESO}.
With repulsion (Fig.~\ref{fig:linstab}), 
the transversal instability region near the $\mu[\rho]=0$ line becomes very thin,
and a new, mostly longitudinal instability 
of the uniform nematic solution emerges in the large density, low noise, and low reversal rate region (Fig.~\ref{fig:linstab}a).

\begin{figure}[t!]
\includegraphics[width=\columnwidth]{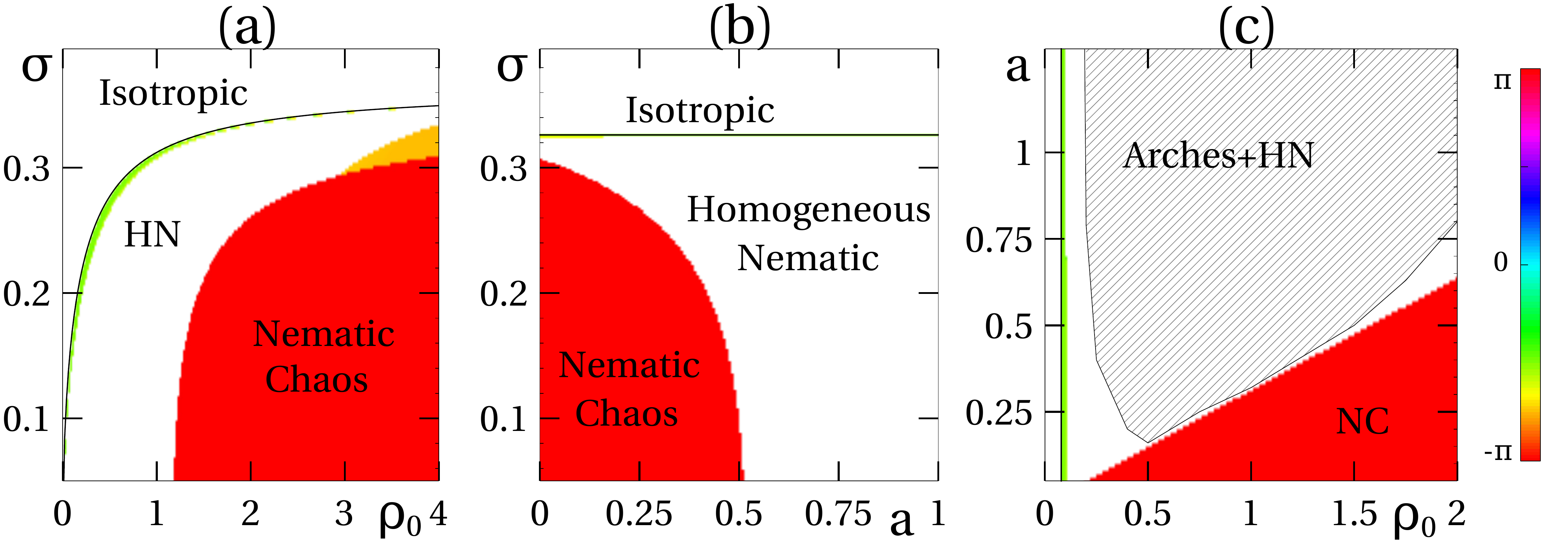}
\caption{Phase diagram of hydrodynamic equations.
(a): $(\rho_0,\sigma)$ plane for $a=0.4$.
(b): $(a,\sigma)$ plane for $\rho_0=1.5$.
(c): $(\rho_0,a)$ plane for $\sigma=0.15$.
Colormap: angle of most unstable wavenumber, if any.
The thin green region along the basic order/disorder line is the transverse banding instability.
Orange region: longitudinal instability leading to nematic chaos. 
In the grey region, multiple arch solutions coexist with the homogeneous nematic state (shown only on panel (c)).
}
\label{fig:linstab}
\end{figure}

We now study the inhomogeneous solutions of our hydrodynamic equations.
As in the microscopic model, the longitudinal instability 
never saturates into stable undulations if the system size is large enough.
It always leads to a nematic chaos regime qualitatively similar to that of the particle model. 
Where only the homogeneous nematic state is stable, we found no other solution, so that the regions 
of linear instability described in Fig.~\ref{fig:linstab} are one to one with those of nematic chaos.

Equations~(\ref{eq:f0},\ref{eq:f1},\ref{eq:f2}) also support stable arch solutions coexisting with the homogeneous nematic state. 
These solutions are qualitatively similar to those observed in the microscopic model (Fig.~\ref{fig:arches}c): 
at given parameter values, arches have a minimal width but no typical/preferred size. They eventually form a regular
smectic pattern. They are globally polar objects with polar drift now encoded explicitly in the $f_1$ field. 
Note however, that the $f_1$ field profiles are rather different from the polar drift profiles recorded in the microscopic model
(Fig.~\ref{fig:arches}c), indicating a strong influence of fluctuations on this typically small field. 
The stability domain of arches is somewhat tedious to determine. A careful numerical investigation for a fixed-size system
revealed that it is only in rough qualitative agreement with that found in the microscopic model (Fig.~\ref{fig:linstab}c).
Again, we attribute this difference to the absence of fluctuations: 
near the bottom of the grey region in Fig.~\ref{fig:linstab}c, arch solutions are ``fragile" and probably do not resist even small amounts of noise.
At the fluctuating level of the particle-based model, all these fragile solutions disappear, leaving the smaller regions of stable arches reported in
Figs.~\ref{fig:micro-phasediagram}c,d.

\begin{figure}
\includegraphics[width=0.95\columnwidth]{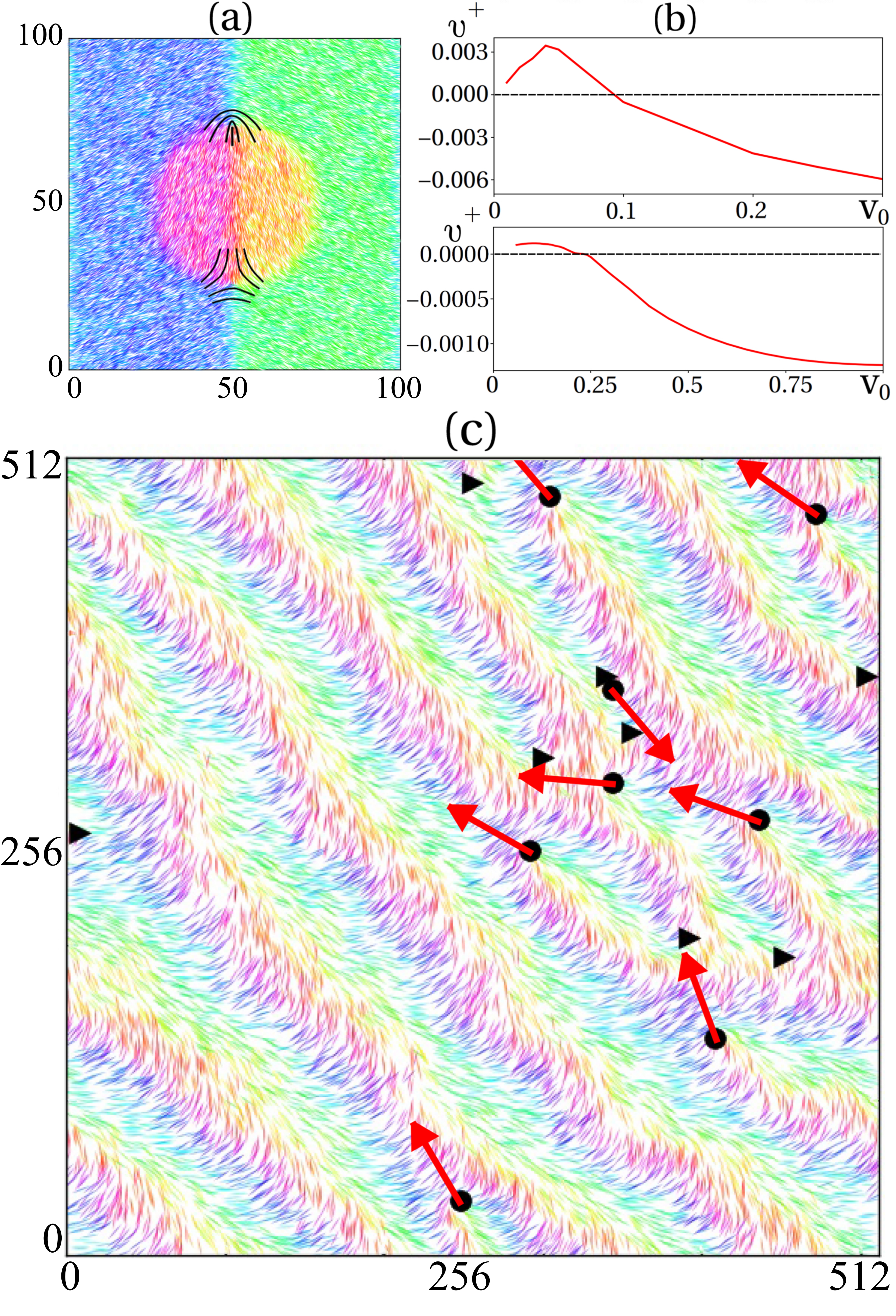}
\caption{Defects dynamics and defect ordered states. 
(a): typical configuration during initial motion at 'asymptotic' velocities 
(microscopic model).
(b): asymptotic velocity $v^+$ of the $+\frac{1}{2}$ defect (along the vertical axis in (a)) vs $v_0$. 
Top: microscopic model ($\rho_0=4$, $\eta=0.1$, $\beta=0.5$, $k=0.5$).
Bottom: hydrodynamic equations ($\rho_0=3$, $\sigma=0.15$, $a=8.0$, $b_1=0.25$)
(c) snapshot of defect ordered state (microscopic model, $\rho_0=4$, $\eta=0.1$, $\beta=0.5$, $k=0.5$). 
Colors as in Fig.~\ref{fig:arches}a with superimposed $-\frac{1}{2}$ (triangles) and $+\frac{1}{2}$ defects (circles with red arrows showing the direction of motion). 
}
\label{fig:defect}
\end{figure}

We now examine in some detail the dynamics and interaction of $\pm\frac{1}{2}$ defects in both our microscopic model and
hydrodynamic equations. Previous works have investigated defects in the nematic chaos (or nematic turbulence) regime, mostly in wet
systems
\cite{SHI-MA-NAT-COMM,YEOMANS-DEFECT,MCM-DEFECT1,MCM-DEFECT2,YEOMANS-VORTICITY,GIOMI-PRX,WET-TO-DRY,BASKARAN,YEOMANS-CELL-DEFECT,DUNKEL,SHANKAR,GENKIN17,GENKIN18,LIVERPOOL18,SELINGER19}. 
Here, in an attempt to disentangle the consequences of the linear longitudinal instability (leading to chaos)
from more intrinsic defect properties, 
we study the fate of a  $\pm\frac{1}{2}$ pair of defects introduced in the uniform
nematic state, using parameters values for which this state is the only stable solution. 
In such conditions, the two defects eventually merge and annihilate, since
the spontaneous nucleation of a pair is never observed on the scales studied here. 
Increasing system size and the initial distance separating defects they can be observed for very long times.
We mostly studied the fate of the configuration shown in Fig.~\ref{fig:defect}a.
For both the microscopic model and the hydrodynamic equations, we first observe that the $+\frac{1}{2}$ defect
has a well-defined asymptotic velocity $v^+$ in the limit of infinite separation distance. 
The $-\frac{1}{2}$ defect, on the other hand, has a vanishing velocity in this limit
with a diffusive behavior driven by the fluctuations. This is in agreement with previous works. 
However, the velocity of the $+\frac{1}{2}$ defect depends strikingly on the nominal particle speed $v_0$ and the reversal rate:
in both the microscopic model and the hydrodynamic equations, 
it can even change sign (Fig.~\ref{fig:defect}b), 
whereas $+\frac{1}{2}$ defects were only reported to move with their cap ahead (i.e. upward here) before.
(Note though that both signs are considered in \cite{SHANKAR,SELINGER19}.)
We conjecture that in wet active nematics
the velocity of the $+\frac{1}{2}$ defect {\it in the fluid frame} could be found to take a different sign depending on 
swimming speed and/or reversal rate of the active particles involved.

This has direct consequences on the existence of defect ordered states.  
No such regime was observed at the deterministic hydrodynamic level.
This remarkable dynamics arises only for the microscopic model when  $v^+>0$,
for parameter values bordering the domain of observation of arch solutions, where
fluctuations are strong enough to nucleate defect pairs.
When $v^+>0$, freshly created pairs unbind quasi-deterministically (for $v^+<0$, they recombine), 
and they continuously remodel the underlying arch pattern without breaking its global polar order (Fig.~\ref{fig:defect}c). 
Thus, the global ordering of $+\frac{1}{2}$ defects ---they flock--- only reflects that of the nearby arch pattern.
This elucidates the origin of the defect ordered state reported for the microscopic model studied in \cite{DOGIC-NATMAT}
(see also \cite{DUNKEL,SRIVASTAVA} for similar, but distinct situations).

To summarize, we have bridged the gap between the microscopic and the hydrodynamic level in dense, dry active nematics.
The Boltzmann-Ginzburg-Landau approach has provided well-behaved hydrodynamic equations 
whose solutions are in good qualitative agreement
with the collective states of the original particle model. This comparison was only possible thanks to
the expression of hydrodynamic transport coefficients in terms of the microscopic control parameters. 
With respect to the dilute case, the phase diagram contains two main new features: (i) a large region at low reversal rate where the homogeneous nematic state is unstable {\it at low noise} and leaves places to nematic chaos, and (ii) multiple arch solutions that coexist with the nematic state at large reversal rate.
These solutions form globally polar smectic patterns at the origin of the heretofore somewhat mysterious defect ordered states. 
We also demonstrated that the properties of the ubiquitous $\pm\frac{1}{2}$ topological defects depend strongly on 
microscopic parameters such as the reversal rate and the nominal speed of particles.

Future work will be devoted to obtaining
a connection between the microscopic and hydrodynamic level similar to that presented here in the case of wet active nematics. 


\end{document}